\documentclass[%
,twocolumn%
,secnumarabic%
,amssymb,aps,pra,nobibnotes]{revtex4-1}
\usepackage{docs}
\usepackage{graphicx}
\usepackage[misc]{ifsym}
\usepackage{subfigure}
\usepackage[left]{lineno}
\usepackage{amsthm,amsmath,amssymb}
\usepackage{mathrsfs}
\usepackage{stfloats}
\usepackage{lipsum}
\usepackage{amsmath,epsfig,amssymb,subfigure,bm,dsfont}
\usepackage{float}

\expandafter\ifx\csname package@font\endcsname\relax\else
 \expandafter\expandafter
 \expandafter\usepackage
 \expandafter\expandafter
 \expandafter{\csname package@font\endcsname}%
\fi
\DeclareRobustCommand\substyle{\name@idx{document substyle}}%
\DeclareRobustCommand\classoption{\name@idx{document class option}}%
\DeclareRobustCommand\classname{\name@idx{document class}}%
\def\name@idx#1#2{%
 {\ttfamily#2}%
 \index{#2\space#1=\string\ttt{#2}\space#1}\index{#1>#2=\string\ttt{#2}}%
}%

\begin{document}
\title{The distribution of B-site in the perovskite for a d$^5$-d$^3$ superexchange system studied with Molecular field theory and Monte Carlo simulation}%
\author{Jiajun Mo$^{1 a}$, Min Liu$^{1a}$, Shiyu Xu$^a$, Qinghang Zhang$^a$, Jiyu Shen$^a$, Puyue Xia$^a$} %
\author{Yanfang Xia$^{a*}$}
\author{Jizhou Jiang$^{b}$}
\thanks{Corresponding author\\ Yanfang Xia: xiayfusc@126.com\\Jizhou Jiang: 027wit@163.com}
\affiliation{a.College of Nuclear Science and Technology, University of South China, Hengyang 421001, China;\\
b.School of Environmental Ecology and Biological Engineering, School of Chemistry and Environmental Engineering, Key Laboratory of Green Chemical Engineering Process of Ministry of Education, Engineering Research Center of Phosphorus Resources Development and Utilization of Ministry of Education, Wuhan Institute of Technology, Wuhan 430205, China.}
\date{February 2022}%
\thanks{the first two authors contributed equally to this work and can be considered co-first authors}
\begin{abstract}
The B-site disorder in the d$^5$ - d$^3$ system of perovskites has been analyzed with molecular field theory and Monte Carlo method. The model is applicable to RFe$_{1-p}$Cr$_p$O$_3$ at any \textit{p} value. When the saturation magnetization $M_S$ and phase transition temperature \textit{T}$_P$ are known, a model can be built to calculate the order or disorder distribution of lattice B-sites. We analyze the case that the Fe-Cr superexchange is antiferromagnetic and ferromagnetic coupling respectively. The simulation result shows that the theoretical calculation formula is suitable for the calculation of different B-site distribution. Through the simulation, we find that when the \textit{x} and \textit{y} are large, the system will appear obvious long-range order. The DM interaction has a certain influence on the saturation magnetization. Via calculation, we found that the distribution states of Fe and Cr do not always conform to the uniform distribution but rather exhibit an effect that reduces the Fe-Fe clustering. The establishment of this model offers an explanation for several previously contentious issues, \textit{e.g.}, what is the phase transition temperature range of double perovskite, and why the different phase transition temperatures with the same doping proportion. It provides theoretical guidance for the design of functional materials with an arbitrary phase transition temperature.\\
\\
\textbf{Keywords}: B-site disorder distribution, Molecular field theory, Heisenberg Model, Monte Carlo
\end{abstract}

\maketitle
\section{Introduction}
Perovskite has the formula RBX$_3$, where R represents nonmagnetic or rare-earth trivalent ions (Bi$^{3+}$, La$^{3+}$, Sm$^{3+}$, Ho$^{3+}$, Gd$^{3+}$, \textit{etc.}), B denotes trivalent transition metal ions. X is a negative divalent anion, most frequently oxygen ion. The B-site d$^5$-d$^3$ system, commonly Fe$^{3+}$ (d$^5$, S = 5/2)-Cr$^{3+}$ (d$^3$, S = 3/2), has a cubic-symmetry ideal structure for RFe$_{1-p}$Cr$_p$O$_3$. In a ideal structure of RFe$_{0.5}$Cr$_{0.5}$O$_3$, Fe$^{3+}$ and Cr$^{3+}$ cations occupy B sites alternately. In this case, the formula can be transformed to R$_2$FeCrO$_6$, a structure known as a double-perovskite. R$_2$FeCrO$_6$, exhibits strong ferromagnetism where the ferromagnetic (FM) coupling exists between Fe$^{3+}$ and Cr$^{3+}$, resulting in a net magnetic moment of 4 $\mu_B$. However, forming the above structure necessitates extremely stringent preparation conditions due to the similar radius of Cr$^{3+}$ and Fe$^{3+}$ ions [1-4]. In reality, the system structure deviates slightly from the ideal structure [5, 6], owing to the disorder of Cr$^{3+}$ and Fe$^{3+}$. It is referred to as anti-site (AS) defects, where Cr$^{3+}$ ions occupy the site previously occupied by Fe$^{3+}$ ions and vice versa. The presence of AS defects significantly reduces the system magnetization. According to the Goodenough-Kanamori (GK) rule [7, 8], Fe$^{3+}$-O-Fe$^{3+}$ and Cr$^{3+}$-O-Cr$^{3+}$ in perovskites are more prone to antiferromagnetic superexchange, and this causes pairwise cancellation of neighboring Fe$^{3+}$/Cr$^{3+}$ moments, significantly reducing the total magnetic moment. In many cases, the saturation magnetic moment of the prepared double perovskite is less than the theoretical magnetic moment; \textit{e.g.} the theoretical saturation magnetic moment of Sr$_2$FeMoO$_6$ of approximately 4 $\mu_B$, and values greater than 3.8 $\mu_B$ have not been reported [9]. This is interpreted as the AS defects of Sr$_2$FeMoO$_6$. However, this phenomenon is not well understood. This work aims to comprehend what AS defect is, starting from, the fundamental model. \par 
In magnetic analysis, the Heisenberg model is a frequently used model. Exchange interaction theory is based on Heisenberg exchange. The Heisenberg model can be used to demonstrate the effects of AS on magnetic properties more intuitively. In most previous reports, a parameter known as anti-site degree (ASD) is used to describe anti-site defect degree. ASD is defined as the ratio of Cr$^{3+}$ ions occupying the site of Fe$^{3+}$ ions to the total amount of  Cr$^{3+}$. We demonstrate in this paper that a single parameter used solely to describe saturation magnetization cannot provide additional information. Indeed, we can approximate the magnetization law of the entire system with only two parameters. The ASD is calculated primarily using the formula $M_S$ = $\mu_{th}$(1-2ASD) [10], where \textit{M}$_S$ is the saturation magnetization and $\mu_{th}$ is the theoretical moment. However, under the antisymmetric anisotropy Dzyaloshinskii-Moriya (DM) interaction [11, 12], antiferromagnetic superexchange possesses a canting moment, conferring the system with weak ferromagnetism (WFM). The DM interaction will affect the ASD calculation in this case. However, the net magnetic moment produced by the DM interaction is often slight, which requires further discussion.\par
 \section{Model establishment}
 \subsection{B-site disorder distribution analysis}
The ideal and anti-site structures are depicted in Fig. 1. We mark the blue site in the ideal structure as the \textit{a}-site and the green as the \textit{b}-site. The Fe$^a$ is the name given to Fe$^{3+}$ placed in \textit{a}-site, and the same is true for Fe$^b$, Cr$^a$ and Cr$^b$. Only Fe$^b$ and Cr$^a$ have an ideal structure, whereas all Fe$^a$, Fe$^b$, Cr$^a$ and Cr$^b$ have an AS defects structure. In this case, the AS defects structure contains four distinct types of nearest superexchange pairs: Fe$^a$-O-Cr$^b$, Fe$^b$-O-Cr$^a$, Fe$^a$-O-Fe$^b$, and Cr$^a$-O-Cr$^b$. Due to the complexity of the AS defect structure, it is necessary to introduce at least three types of exchange constants, namely \textit{J}$_{FC}$,\textit{ J}$_{FF}$ and \textit{J}$_{CC}$, which represent exchange coupling between Fe$^{3+}$ and Cr$^{3+}$, Fe$^{3+}$ and Fe$^{3+}$, Cr$^{3+}$ and Cr$^{3+}$, respectively. In the following, all the $\mathrm{'}$Fe$\mathrm{'}$ are abbreviated to $'$F$'$, the same as $'$Cr$'$. The complex magnetic system is based on the interaction of a single site with its adjacent sites. The study of isolated sites is complex and frequently necessitates specialized processing methods. Classical molecular field theory (MFT) is the best solution to approximate the complex magnetic system, typically dividing the crystal into two or more sublattices. This paper studies four sublattices, \textit{i.e.}, F$^{a}$, F$^b$, C$^a$, and C$^b$. 
\begin{figure}[H]
    \centering
    \includegraphics[scale=0.361]{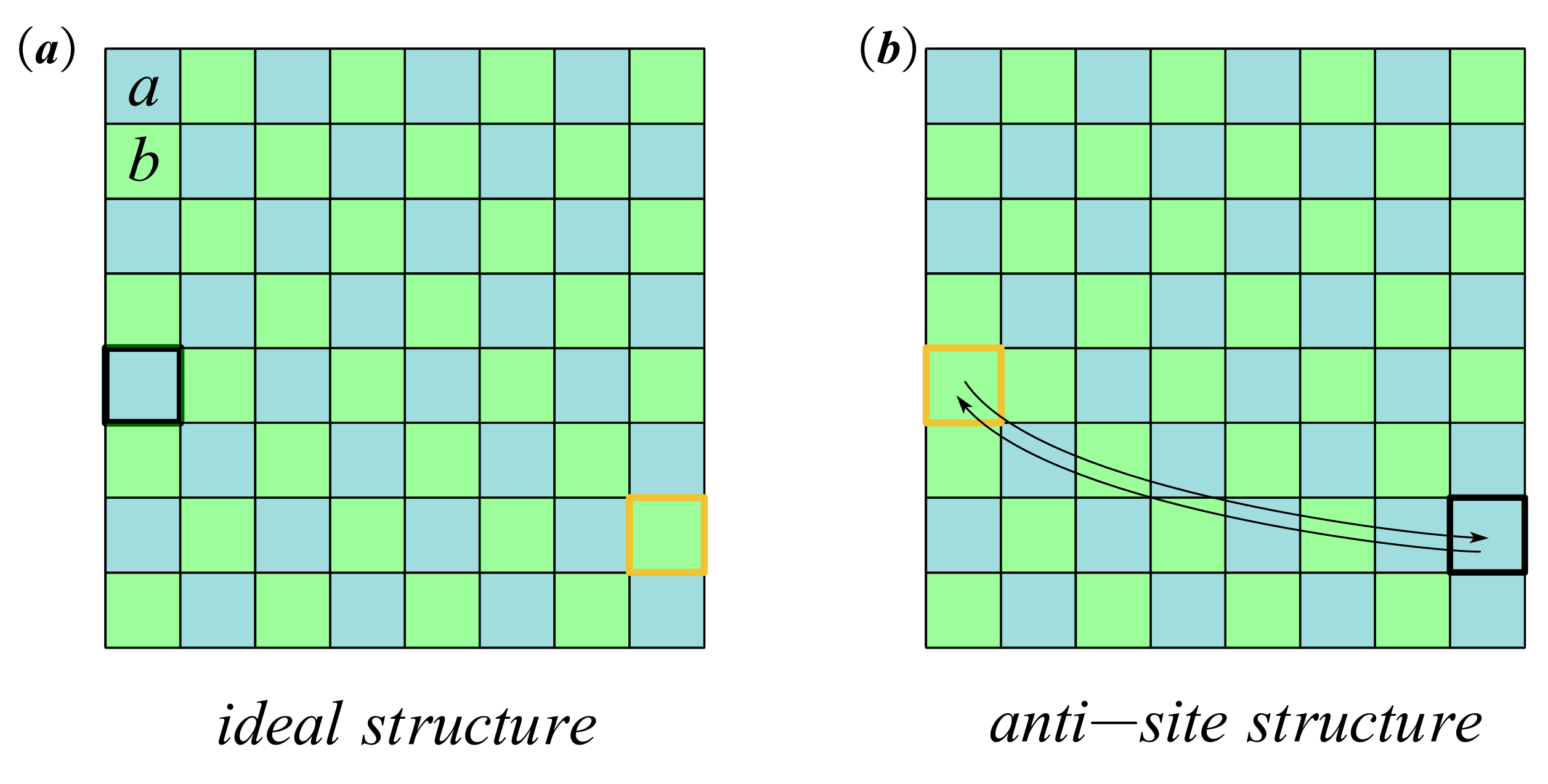}
    \caption{The plot of ideal structure and anti-site structure, the blue and green squares represent Cr$^{3+}$ and Fe$^{3+}$, respectively.}
\end{figure}
Let \textit{Z}$_{ij}$ denote the average number of \textit{j}-site surrounding \textit{i}-site, \textit{e.g.}, in an ideal structure, Z$_{{F}^b{C}^a}$ = \textit{Z}$_{{C}^a{F}^b}$ = 6. Following the preceding analysis, we can derive the correlating relationship (in AS defects structure) as $Z_{i^aj^b}=Z-Z_{i^ai^b}$, where \textit{Z} is the total number of coordination between B-site ions(\textit{Z} equals six in perovskite). \textit{i} and \textit{j} represent arbitrary one of Fe and Cr respectively, and \textit{i} and \textit{j} are different. Considering the conservation of total exchange pairs, \textit{i.e.}, the number of Fe$^a$-O-Fe$^b$ pairs is equal to that of Fe$^b$-O-Fe$^a$, namely
\begin{equation}
    n_iZ_{ij}=n_jZ_{ji},
\end{equation}
wherein, \textit{n}$_i$ represents the molar fraction of the \textit{i}-ion. We define a more general parameter \textit{f}, which is applicable to any \textit{p} in RFe$_{1-p}$Cr$_p$O$_3$ to describe AS defect, as opposed to ASD which is applicable only to double perovskite. The \textit{f} factor can be defined as
\begin{equation}
    f=\begin{cases}
	\frac{n_{C^b}}{n_{C^a}+n_{C^b}}\,\,  0<p\le 0.5\\
	\frac{n_{F^b}}{n_{F^a}+n_{F^b}}\,\,  0.5<p<1\\
\end{cases}\,\,
\end{equation}
According to Eq. 2 and the formulas $n_{C^a}+n_{C^b}=p, n{_{F^a}}+n_{F^b} = 1-p$, it can be calculated that 
\begin{gather}
n_{C^b}=pf,\, n_{C^a}=p\left( 1-f \right),\nonumber\\
n_{F^b}=0.5-pf,\, n_{F^a}=0.5-p\left( 1-f \right),
\end{gather}
when \textit{p} is less than or equal to 0.5. By combining formula (1), (2), and (3), we can obtain the relation between \textit{f} and Z$_{F^aF^b}$, Z$_{F^bF^a}$, as well as that between Z$_{F^aF^b}$ (abbreviated as \textit{x}), Z$_{F^bF^a}$ (abbreviated as \textit{y}) and Z$_{C^aF^b}$, Z$_{C^bF^a}$.
\begin{equation}
    f=\frac{0.5y-\left( 0.5-p \right) x}{p\left( x+y \right)}
\end{equation}
\begin{footnotesize}
\begin{gather}
    Z_{C^aF^b}=6-{Z_{C^aC^b}}=\frac{2x\left( 1-p \right) \left( 6-y \right)}{x-y+2py}\nonumber\\
    Z_{C^bF^a}=6-Z_{C^bC^a}=\frac{2y\left( 1-p \right) \left( 6-x \right)}{y-x+2px}
\end{gather}
\end{footnotesize}
Until now, all \textit{Z}$_{ij}$ and \textit{f} factor have been expressed in terms of two parameters, \textit{x} and \textit{y}. Additionally, as the value of Z$_{ij}$ is limited between 0 and 6, and the value of  \textit{f} is limited between 0 and 1, the range of values for \textit{x} and \textit{y} are determined, that is,
\begin{small}
\begin{equation}
    s.t. \left\{ \begin{array}{c}
	\begin{array}{l}
	x\left( 1-p \right) \left( 6-y \right) \le 3\left( x-y+2py \right)\\
	y\left( 1-p \right) \left( 6-x \right) \le 3\left( y-x+2px \right)\\
\end{array}\\
	0\le x\le 6\\
	0\le y\le 6\nonumber\\
\end{array} \right. 
\end{equation}
\end{small}
within the above constraints, \textit{x} and \textit{y} can be adjusted arbitrarily.\par
\begin{figure}[htbp]
    \centering
    \includegraphics[scale=0.42]{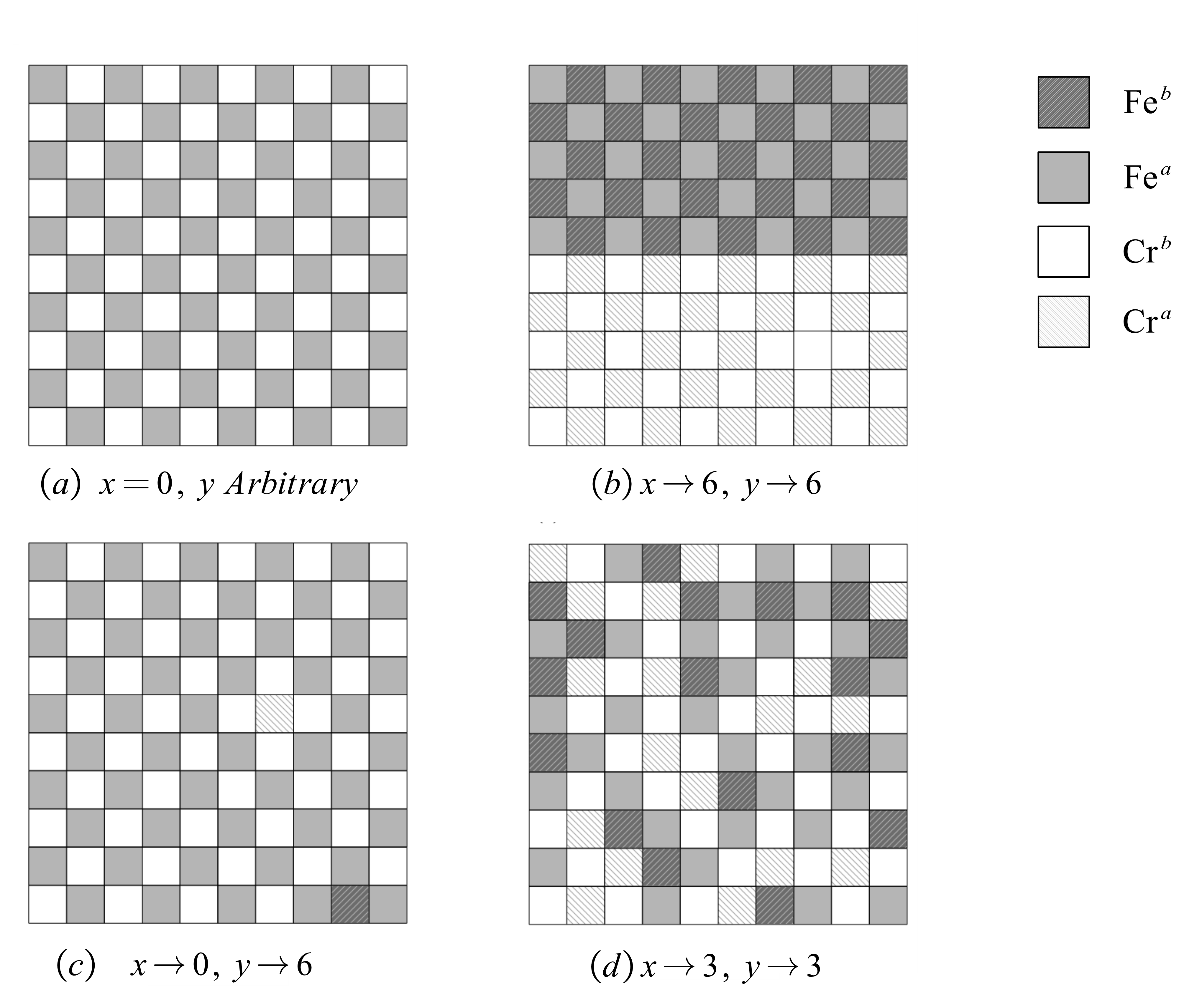}
    \caption{The arrangement diagram of anti-site atoms. (a) is spaced when \textit{x} = \textit{y} = 0, (b) is entirely separate, Fe and Cr atoms occupy both sides to form Fe-rich and Cr-rich. (c) there is anti-site of very few atoms in the spaced arrangement. If the anti-site atoms are not clustered, \textit{y} will tend to be 6. (d) is an entirely random chaotic state. \textit{x} and \textit{y} tend to 3 with the increase of the number of atoms.}
\end{figure}
The ideal double perovskite structure is shown in Fig. 2 (a), where \textit{y} can be any value. When $x\rightarrow$ 6 and $y\rightarrow$ 6, the Fe-rich and Cr-rich phenomena are evident. One could even call that the system contains two independent phases with Fe$^{3+}$ and Cr$^{3+}$ occupying one side, as illustrated in Fig. 2(b). $x\rightarrow$ 0, $y\rightarrow$ 6 is depicted as in Fig. 2(c), the anti-site atoms essentially do not exist, \textit{i.e.}, it can be ignored Fe$^{3+}$-O-Fe$^{3+}$ or Cr$^{3+}$-O-Cr$^{3+}$ exchange. When $x \rightarrow$ 3 and $y \rightarrow$ 3, the arrangement tends to disorder (Fig. 2(d)), this situation corresponds to the case of uniform distribution. The preceding rules can be summarized in the following manner. When \textit{x} and \textit{y} are both large ($x\rightarrow$ 6, $y\rightarrow$ 6), they are predominantly highly clustered, \textit{i.e.} there are obvious dislocation clusters, which account for a sizable proportion. The stronger the independence between Fe and Cr, the more they form clusters. When one of \textit{x} and \textit{y} is small and the other is large, it indicates that anti-site atoms are dispersed and there is no apparent agglomeration of cluster. When \textit{x} and \textit{y} are small, it indicates that the whole system roughly is in the form of orderly interval arrangement, and less proportion of dislocation atoms.
 \subsection{Construct the disordered distribution sites and establish Heisenberg model}
 We use a simple method to construct any site distribution corresponding to \textit{x} and \textit{y}:\\
\textbf{Step 1}, construct \textit{L}$\times$\textit{L}$\times$\textit{L} lattice, randomly arrange the positions of Cr and Fe according to the set proportion \textit{p}.\\
\textbf{Step 2}, calculate each coordination number $Z_{ij}$ under the initial random.\\
\textbf{Step 3}, calculate the difference $\rho$ between $Z_{F^aF^b}$, $Z_{F^bF^a}$ and \textit{x}, \textit{y}, \textit{i.e.}, $\rho =|Z_{F^aF^b}-x|+|Z_{F^bF^a}-y|$\\
\textbf{Step 4}, randomly select a Cr, if it is in \textit{a}-site, exchange it with a random Fe$^b$ site, then calculate the difference $\rho'$ between Z$'_{F^aF^b}$, Z$'_{F^bF^a}$ and \textit{x}. If $\rho'$ $<$ $\rho$, accept for this exchange, otherwise reject this exchange and return to original position.\\
\textbf{Step 5}, repeat \textbf{Step 4} until $\rho$ is 0.
Go through the above five steps to obtain the lattice distribution of the specified coordination.\par
In order to investigate the effect of disorder on the magnetic properties of perovskite, we will consider DM interaction. The Hamiltonian of the system can be expressed as
\begin{equation}
    \mathcal{H} =-\sum_{\left< i,j \right>}{J_{ij}\mathbf{S}_i\mathbf{S}_j}-\sum_{\left< i,j \right>}{\mathbf{D}_{ij}·\left[ \mathbf{S}_i\times \mathbf{S}_j \right]}\nonumber
\end{equation}
wherein, $S_i$ is the pseudospin vector and let $S_{F}$= 5/2, $S_{C}$ = 3/2. $\left< i,j \right> $ represents the nearest neighbour of \textit{i} is \textit{j} ion, the $J_{ij}$ is the exchange constant between \textit{i} and \textit{j} ions. $D_{ij}$ is the DM interaction vector. There is a cross product in the above Hamiltonian. In order to simplify the calculation, we only study the DM interaction in the \textit{z}-axis. So, the Hamiltonian can be rewritten as 
 \begin{equation}
     \mathcal{H} =-\sum_{\left< i,j \right>}{J_{ij}\mathbf{S}_i\mathbf{S}_j}-\sum_{\left< i,j \right>}{D_{ij}^{}·\left( S_{i}^{x}S_{j}^{y}-S_{i}^{y}S_{j}^{x} \right)}
 \end{equation}
The $D_{ij}$ in formula (6) is an antisymmetric parameter, \textit{i.e.}, $D_{ij}$ = -$D_{ji}$.  In this work, we let $|D_{FC}|$ = $|D_{CC}|$ = $|D_{FF}|$ = \textit{D} to reduce the number of parameters.
\subsection{Analyze the relationship between $T_P$, $M_S$ and parameters \textit{x}, \textit{y}}

We establish four sublattices to calculate the phase transition temperature, ignoring the effect of DM interaction to reduce computational complexity. Each sublattice has an energy value:
\begin{gather}
        H_{F^a}=\lambda _{F^aF^b}M_{F^b}+\lambda _{F^aC^b}M_{C^b}+H\notag\\
        H_{F^b}=\lambda_{F^bF^a}M_{F^a}+\lambda_{F^bC^a}M_{C^a}+H\\
        H_{F^b}=\lambda _{F^bF^a}M_{F^a}+\lambda _{F^bC^a}M_{C^a}+H\notag\\
        H_{C^a}={\lambda _{C^a}}_{F^b}M_{F^b}+\lambda _{C^aC^b}M_{C^b}+H\notag
\end{gather}
where $\lambda_{ij}$ =$2Z_{ij}J_{ij}/n_iN_A(g\mu_B)^2$ is the molecular field coefficient of \textit{j}-sublattice relative to \textit{i}-sublattice. Here $J_{ij}$ denotes the exchange constant between \textit{i}- and \textit{j}-ions, \textit{g} is the lande factor, $\mu_B$ represent Bohr magneton. \textit{H} is the external magnetic field. $\textit{M}_i$ represents the magnetization of \textit{i}-sublattice. The magnetization of \textit{i}-sublattice $M_i$ is 
\begin{equation}
    M_i=n_iN_Ag\mu _BS_iB_{S_i}\left( \frac{g\mu _BS_iH_i}{k_BT} \right) 
\end{equation}
where \textit{B}$_{J}(\gamma)$ is the Brillouin function. Near the phase transition temperature $\gamma \ll 1$. The Brillouin function can be reduced to the first term of Taylor expansion as $B_J(\gamma)\approx\gamma(J+1)/3J$. Let \textit{H} = 0, simultaneous equation (6) (7) and (8), solve the relation of $M_i$ and include the coefficient of $M_i$ in the determinant \textit{A}. Let the determinant \textit{A} be 0, \textit{i.e.},
\begin{small}
 \begin{widetext}
\begin{eqnarray}
    A=\left| \begin{matrix}
	-3k_{\mathrm{B}}T_P\,&		2J_{FF}\,S_F\,Z_{F^{\mathrm{a}}F^{\mathrm{b}}}\left( S_F+1 \right)&		0&		2J_{FC}S_FZ_{F^aC^b}\left( S_F+1 \right)\\
	2J_{FF}\,S_F\,Z_{F^{\mathrm{b}}F^{\mathrm{a}}}\left( S_F+1 \right)&		-3k_{\mathrm{B}}T_P&		2J_{FC}\,S_FZ_{F^bC^a}\left( S_F+1 \right)&		0\\
	0&		2J_{FC}\,S_CZ_{C^aF^b}\left( S_C+1 \right)&		-3k_{\mathrm{B}}T_P&		2J_{CC}S_CZ_{C^aC^b}\left( S_C+1 \right)\\
	2J_{FC}\,S_C\,Z_{C^{\mathrm{b}}F^{\mathrm{a}}}\left( S_C+1 \right)&		0&		2J_{CC}S_CZ_{C^bC^a}\left( S_C+1 \right)&		-3k_{\mathrm{B}}T_P\\
\end{matrix} \right|=0
\end{eqnarray}
\end{widetext}
\end{small}
The solution \textit{T}$_P$ of equation (9) is 
\begin{gather}
T_P=\frac{1}{3\sqrt{2}k_B}\sqrt{\begin{array}{l}
	A_{12}A_{21}+A_{14}A_{41}+A_{23}A_{32}+A_{34}A_{43}\nonumber\\
\end{array}}
\\
\,\,      \overline{+\sqrt{\left( A_{12}A_{21}+A_{14}A_{41}+A_{23}A_{32}+A_{34}A_{43} \right) ^2}}\nonumber
\\
\,\,        \overline{\overline{-4\left( A_{21}A_{43}-A_{41}A_{23} \right) \left( A_{12}A_{34}-A_{14}A_{32} \right) }}
\end{gather}
where \textit{A}$_{ij}$ is the item of row \textit{i} and column \textit{j} of the determinant \textit{A}.
To calculate the saturation magnetization in the ground state with DM interaction, the direction of the magnetic moment of each sublattice needs to be considered first. The canted angle is defined with the smallest energy. All spins on each sublattice should point in the same direction in the ground state. In the \textit{x}o\textit{y} plane, the energy \textit{E} per site for the system is
\begin{small}
\begin{gather}
\frac{E}{N}=k_1\left( J_{FF}\cos \alpha _{FF}+D_{FF}\sin \alpha _{FF} \right) +\left( k_2+k_3 \right) J_{FC}\cos \alpha _{FC}\nonumber
\\
+\left( k_2+k_3 \right) D_{FC}\sin \alpha _{FC}+k_4\left( J_{CC}\cos \alpha _{CC}+D_{CC}\sin \alpha _{CC} \right) 
\end{gather}
\end{small}
where
\begin{small}
\begin{gather}
 k_1=Z_{F^aF^b}n_{Fe^a}S_{Fe}^{2}, \,k_2=Z_{F^aC^b}n_{Fe^a}S_FS_C,\nonumber
\\
k_3=Z_{C^aF^b}n_{Cr^a}S_{F}^{}S_{C}^{},\, k_4=Z_{C^aC^b}n_{C^a}S_{C}^{2}\nonumber
\end{gather}
\end{small}
$\alpha_{FF}$, $\alpha_{FC}$, $\alpha_{CC}$ satisfy relation $\alpha _{FC}=\frac{\alpha _{FF}+\alpha _{CC}}{2}$, see in Fig. 3. In order to solve the canted angles, three canted angles that minimize the Hamiltonian must be found [13], namely, let 
\begin{gather}
    \frac{\partial \textit{E}}{\partial \alpha _{FF}}|_{\alpha _{FF}=\alpha _1}=0,\,\,\,\,
\frac{\partial\textit{E}}{\partial \alpha _{CC}}|_{\alpha _{CC}=\alpha _3}=0\nonumber
\end{gather}

It is well established that the canted angle caused by DM interaction is extremely small. We use the Taylor approximation to solve the above $\alpha_{FF}$, $\alpha_{FC}$, $\alpha_{CC}$, putting $\theta_{ij}$ = $\pi-\alpha_{ij}$ into Eq. 11, and taking $\cos \theta _{ij}\approx 1-\frac{{\theta _{ij}}^2}{2}$ ,  $\sin \theta _{ij}\approx \theta _{ij}$. To a solution of 
\begin{footnotesize}
\begin{gather}
\alpha _{FF}=\frac{D_{FF}\,J_{FC}k_1\left( \,k_2+k_3 \right) -D_{CC}\,J_{FC}\,k_4\left( k_2+k_3 \right)}{J_{CC}J_{FC}k_4\left( k_2+k_3 \right) +4J_{CC}J_{FF}\,k_1\,k_4+J_{FC}J_{FF}k_1\left( k_2+k_3 \right)}\nonumber
\\
+\frac{2D_{FC}\,J_{CC}k_4\left( k_2+k_3 \right) +4D_{FF}\,J_{CC}\,k_1\,k_4}{J_{CC}J_{FC}k_4\left( k_2+k_3 \right) +4J_{CC}J_{FF}\,k_1\,k_4+J_{FC}J_{FF}k_1\left( k_2+k_3 \right)}+\pi\nonumber
\\
\alpha _{CC}=\frac{D_{CC}J_{FC}k_4\left( k_2+k_3 \right) -D_{FF}J_{FC}k_1\left( k_2+k_3 \right)}{J_{CC}J_{FC}k_4\left( k_2+k_3 \right) +4J_{CC}J_{FF}k_1\,k_4+J_{FC}J_{FF}k_1\left( k_2+k_3 \right)}\nonumber
\\
+\frac{2D_{FC}\,J_{FF}\,k_{1}^{}\left( k_2+k_3 \right) +4D_{CC}J_{FF}\,k_1\,k_4}{J_{CC}J_{FC}k_4\left( k_2+k_3 \right) +4J_{CC}J_{FF}k_1\,k_4+J_{FC}J_{FF}k_1\left( k_2+k_3 \right)}+\pi\nonumber
\end{gather}
\end{footnotesize}
\begin{figure}[H]
    \centering
    \includegraphics[scale=1]{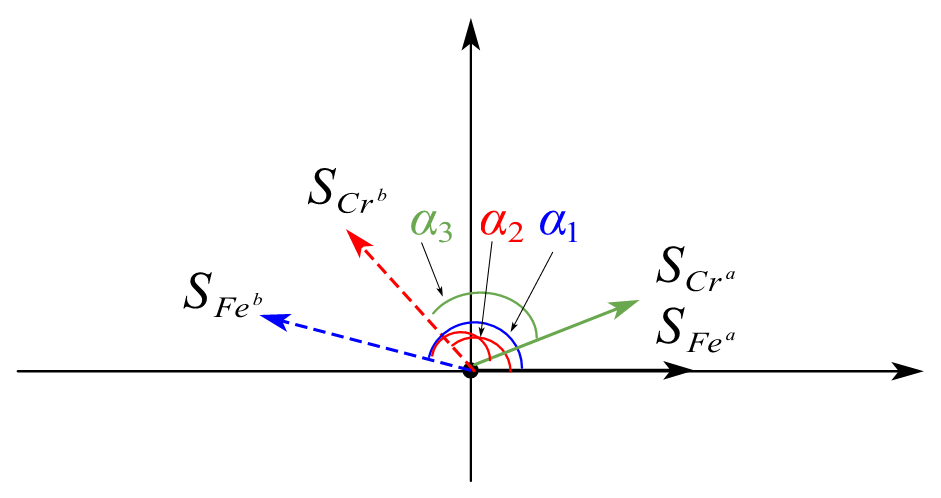}
    \caption{Schematic diagram of competition between two sublattices with different canting angles for DM interaction on the same site. }
\end{figure}
The net magnetic moment can be calculated by the superposition of plane coordinate vectors, as illustrated in Fig. 3. Let $M_{F^a}$: (\textit{n}$_{F^a}{S_F}$,0), $M_{F^b}$:($n_{F^b}S_F$·cos$\alpha_{FF}$, $n_{F^b}S_F$·sin$\alpha_{FF}$), $M_{C^b}$:($n_{C^b}S_C$·cos$\alpha_{FC}$, $n_{C^b}S_C$·sin$\alpha_{FC}$), $M_{C^a}$:($n_{C^a}S_C$·cos($\alpha_{FF}$ - $\alpha_{FC}$), $n_{C^a}S_C$·sin($\alpha_{FF}$ - $\alpha_{FC}$)). The net moment is 
\begin{equation}
M_{net}=\sqrt{A_1+A_2}
\end{equation}
where
\begin{align*}
A_1=&[ n_{F^a}S_F+n_{F^b}S_F\cos \alpha _{FF}\\
&+n_{C^b}S_C\cos \alpha_{FC}+n_{C^a}S_C\cos \left( \alpha _{FF}-\alpha _{FC} \right) ]^2 \nonumber\\
A_2=&[ n_{F^b}S_F\sin \alpha _{FF}+n_{C^b}S_C\sin \alpha _{FC}\\&+n_{C^a}S_C\sin \left( \alpha _{FF}-\alpha _{FC} \right) ]^2 \nonumber
\end{align*}
The above calculation is only applicable to the analysis of Fe-O-Cr antiferromagnetic superexchange. To verify the accuracy of the above theory, we use Monte Carlo method to simulate different coordination \textit{x}, \textit{y} and proportions \textit{p}.
\subsection{Monte Carlo simulation}
We simulate the Heisenberg model using Monte Carlo method based on the Metropolis Hastings criterion. To minimize the time overhead, we optimize the calculation using the parallel chessboard algorithm [14] and Gaussian adaptive sampling [15]. Set \textit{L} equal to 20, as there is no discernible difference between \textit{L} values greater than 20 [16]. According to the Metropolis Hastings criterion, the flip probability \textit{P} is
\begin{equation}
    P=e^{-\frac{\Delta H}{k_BT}}
\end{equation}
where $k_B$ is Boltzmann constant and $\Delta H$ is the energy difference between the per- and post-flip states. \textit{M} should be calculated as
\begin{equation}
    M=\sqrt{m_{x}^{2}+m_{y}^{2}+m_{z}^{2}}
\end{equation}
wherein
$$
m_{\alpha}=\frac{1}{L^3}\sum_{i=1}^{L^3}{S_{i}^{\alpha}}, \left( \alpha =x, y, z \right) 
$$
The susceptibility $\chi$ of this system is a key parameter, which reflect the phase transition, that is described as 
\begin{equation}
\chi =\frac{\partial M}{\partial H}=\frac{L^3\left( \Delta M \right) ^2}{k_BT}=L^3\frac{\left< M^2 \right> -\left< M \right> ^2}{k_BT}
\end{equation}
where \textit{i} represents the \textit{i}-th thermodynamic statistical time. The entire simulation process requires performing flip judgments in a loop; when the loop run for $L^3$ time, the entire process is referred to as one Monte Carlo step (MCs). We use 10$^4$ MCs to bring the system to equilibrium, and 10$^5$ MCs to calculate statistics. \par
\section{Result and Discussion}
Figure 4 illustrates the situation of two orbits coupling, in which, $p\theta$ orbits coupling is the leading factor in Fe-O-Cr FM superexchange. When the bond angle $\theta$ is between 142°$<$ $\theta$ $<$ 156° [17, 18], the orbital coupling transforms as $p\pi$, and the Fe-O-Cr coupling transforms to AF superexchange. The sign of the Fe-O-Cr superexchange coupling constant has been controversial. This superexchange coupling constant is different in many materials. We will avoid debating this point, and instead, study AS in two situations. The $J_{CC}$ is negative sign according to the Goodenough-Kanamori (GK) [7,8], the same holds true for $J_{FF}$. To be consistent with reality, we set $J_{FF}$ = 1.5$J_{CC}$ and $|J_{FC}|$ = 0.5$|J_{CC}|$ to maintain the isometric relationship between the three exchange constants.\par
\begin{figure}[H]
    \centering
    \includegraphics[scale=0.31]{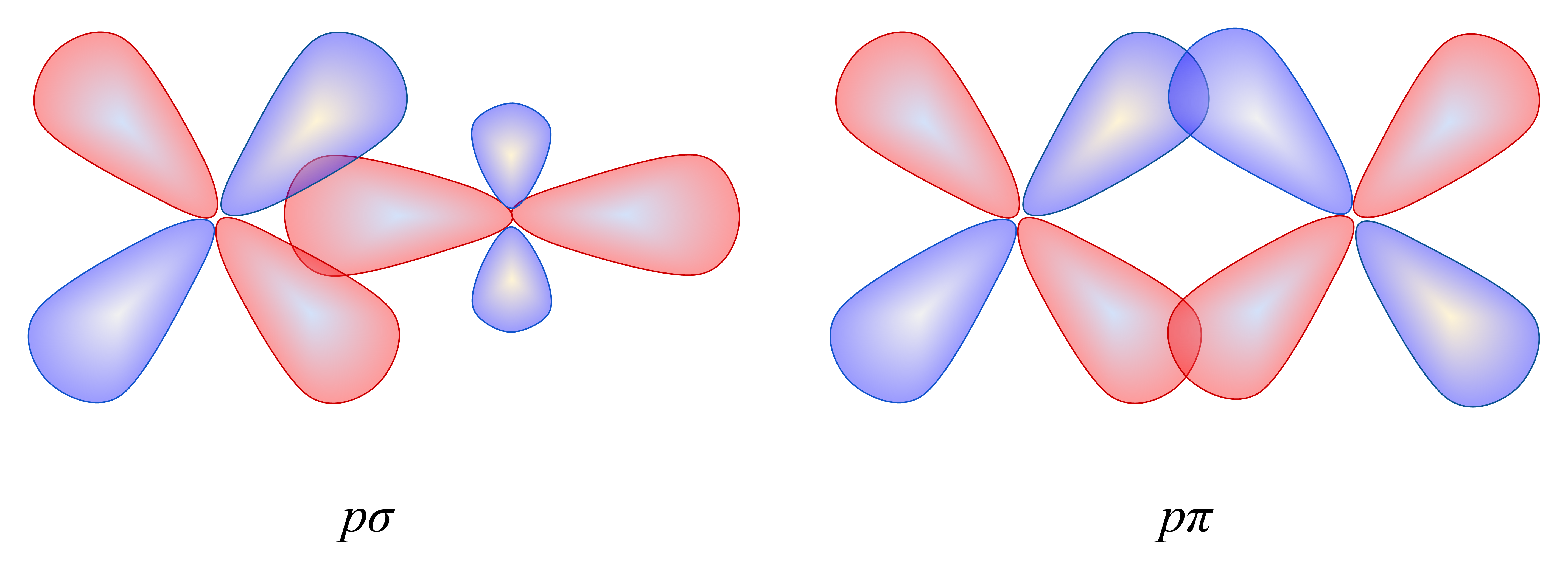}
    \caption{Schematic diagram of the two kinds of orbits coupling}
\end{figure}
For the magnetization curve under various coordination relations, see Fig. 5. When \textit{f} equals 0.4, the saturation magnetization is approximately 0.1, which is attributed to the AF superexchange coupling between Fe and Cr. In any case, all ions in \textit{a}-site always point in the same direction, the same is true of \textit{b}-site. When \textit{f} is equal to 0.4, 40$\%$ of Cr occupies the \textit{b}-site and 60$\%$ Cr occupies the \textit{a}-site. The total moment is easily calculated to be 0.625 in the \textit{a}-site, and 0.525 in the \textit{b}-site. The difference between the moments of \textit{a}- and \textit{b}-site is the net moment. Therefore, the saturation magnetization $M_S$ in RFe$_{0.5}$Cr$_{0.5}$O$_3$ can be summarized as follows:
\begin{equation}
    M_S=|\frac{1-2f}{2}|
\end{equation}
The inset of Figure 5 (a) shows the general rule, which states that the greater the \textit{y}, the higher the phase transformation temperature. \par
\begin{figure}[H]
    \centering
    \includegraphics[scale=0.31]{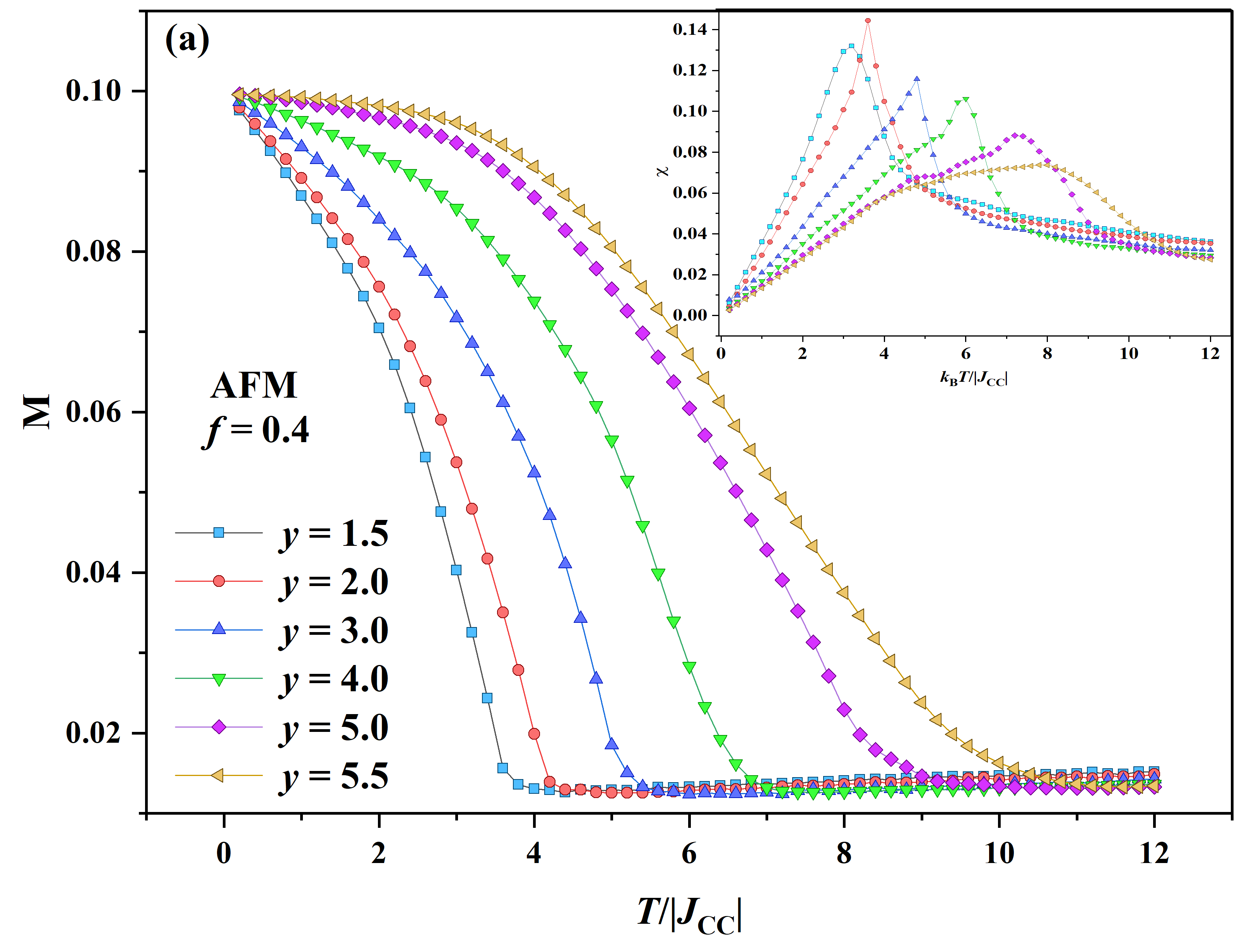}
    \hspace{1in}
    \includegraphics[scale=0.31]{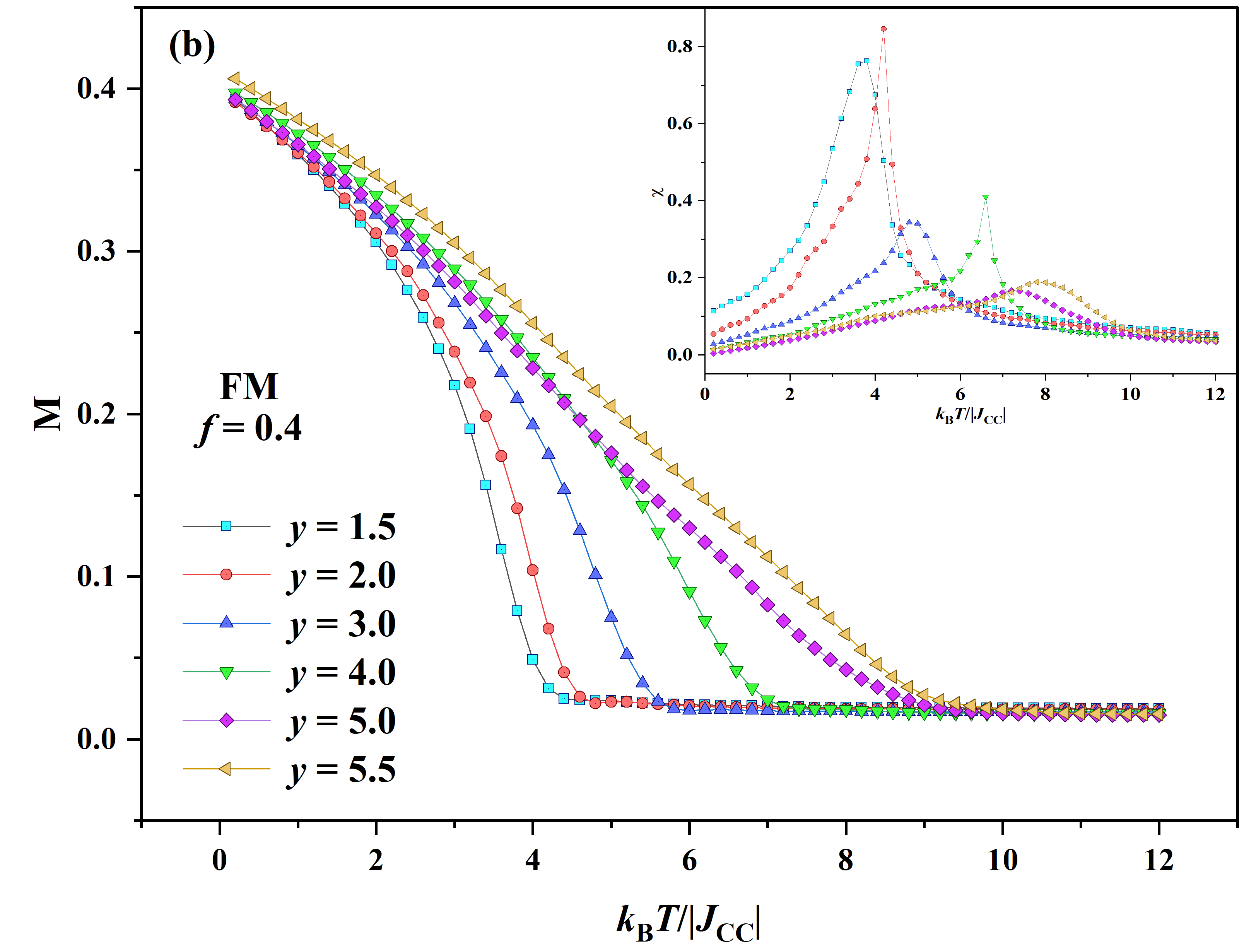}
    \caption{Temperature dependence of magnetization under the same ASD and different coordination number. (a) in the case of the AF superexchange coupling between Fe and Cr, \textit{i.e.}, $J_{FC}$ $<$ 0. (b) in the case of the FM superexchange coupling between Fe and Cr, \textit{i.e.}, $J_{FC}$ $>$ 0. Insets are the susceptibility with different y, which correlating to the M-T.}
\end{figure}
Figure 5 (b) shows the temperature dependence of magnetization under FM superexchange coupling between Fe and Cr. The saturation magnetization is approximately 0.4, which is in the presence of exchange competition. However, there is a dominant tendency in this competitive relationship when the degree of dislocation is small. If \textit{f} = 0, the magnetic moments of the a- and b-site point in the same direction. When \textit{f} = 0.4, 40$\%$ Fe$^b$ takes the place of Cr$^a$. After replacement, the magnetic moment is oriented in the opposite direction to the original magnetic moment orientation. Because the competitive trend dominated by antiferromagnetic interaction maintains the opposite magnetic moment orientation, as shown in Fig. 6. From this, it is calculated that the moment on \textit{a}- and \textit{b}-site is roughly  \begin{gather}
    a-\mathrm{site}:           M^a=\frac{3}{4}\left( 1-f \right) -\frac{5}{4}f=\frac{3}{4}-2f\nonumber
\\
b-\mathrm{site}:           M^b=\frac{5}{4}\left( 1-f \right) -\frac{3}{4}f=\frac{5}{4}-2f\nonumber
\end{gather}
The saturation magnetization $M_S$ is sum of $M_a$ and $M_b$, \textit{i.e.},
\begin{equation}
    M_S=2-4f
\end{equation}
\begin{figure}[H]
    \centering
    \includegraphics[scale=0.31]{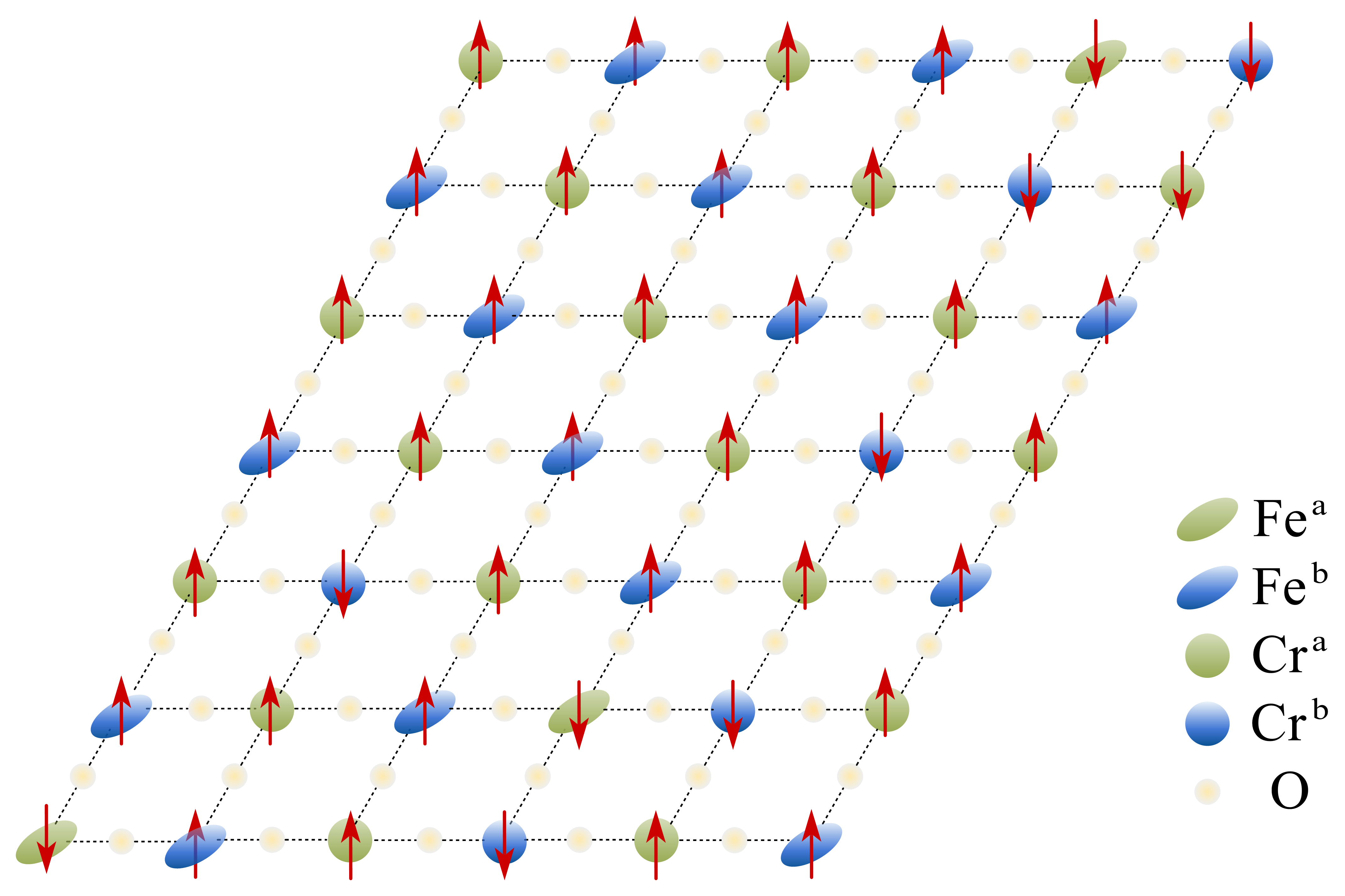}
    \caption{Lattice distribution with magnetic moment orientation in the presence of AS defects.}
\end{figure}
Formula (16),(17) consistent with the formula in Ref. 10 and the formula (12) when \textit{D} = 0. \par FM superexchange coupling uses a phase transformation mechanism similar to that of AF superexchange coupling between Fe and Cr. It is worth noting that when \textit{y} is less than or equal to 4, the magnetic susceptibility peak value exhibits some deviation. While this appears to contradict molecular field theory, it is merely an illusion of frustrated magnetic susceptibility. In the frustrated state, the magnetic susceptibility peak temperature may deviate from the phase transition temperature, which must be determined precisely using heat capacity or fourth-order Binder cumulants. 
\begin{figure}[H]
    \centering
    \includegraphics[scale=0.31]{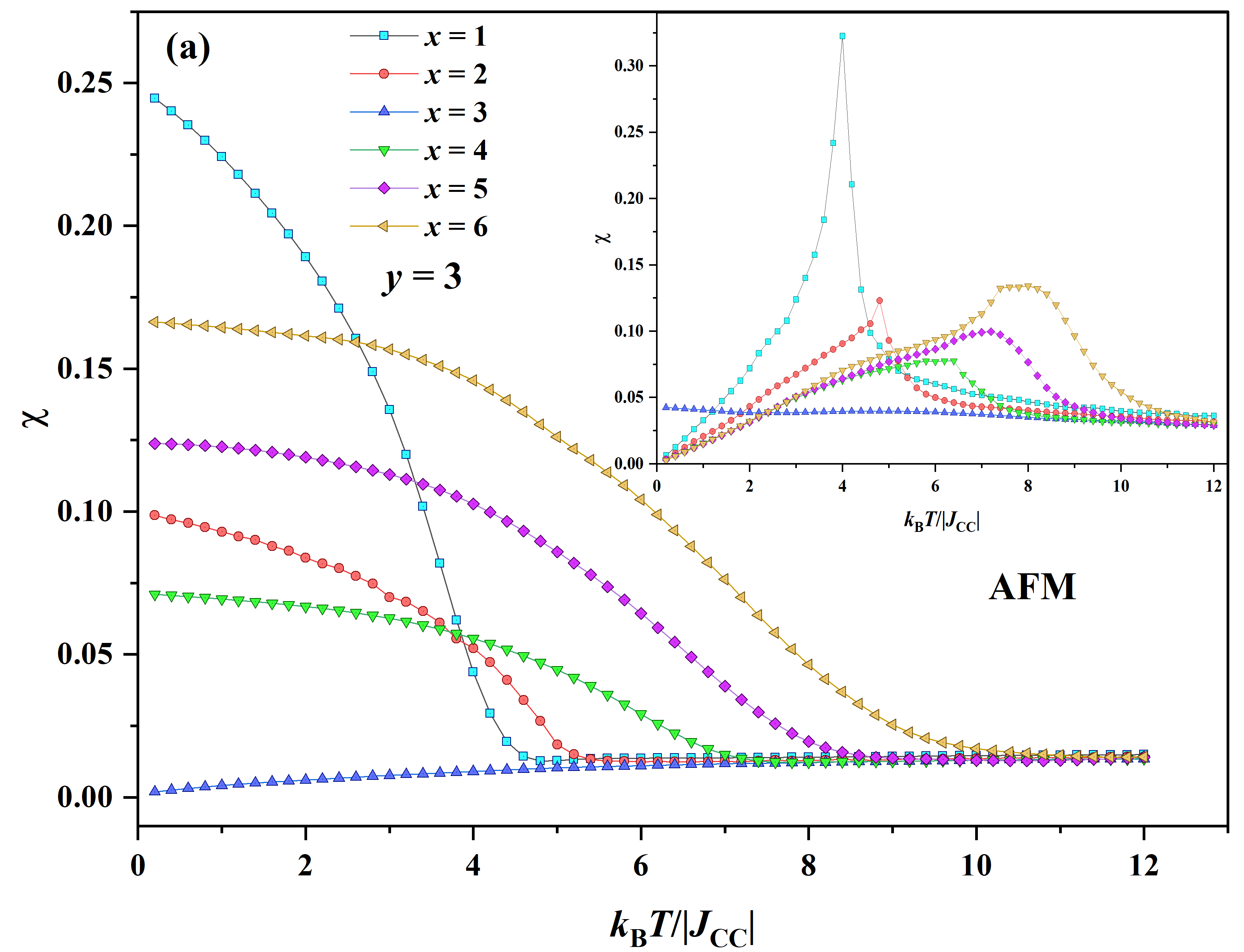}
    \includegraphics[scale=0.31]{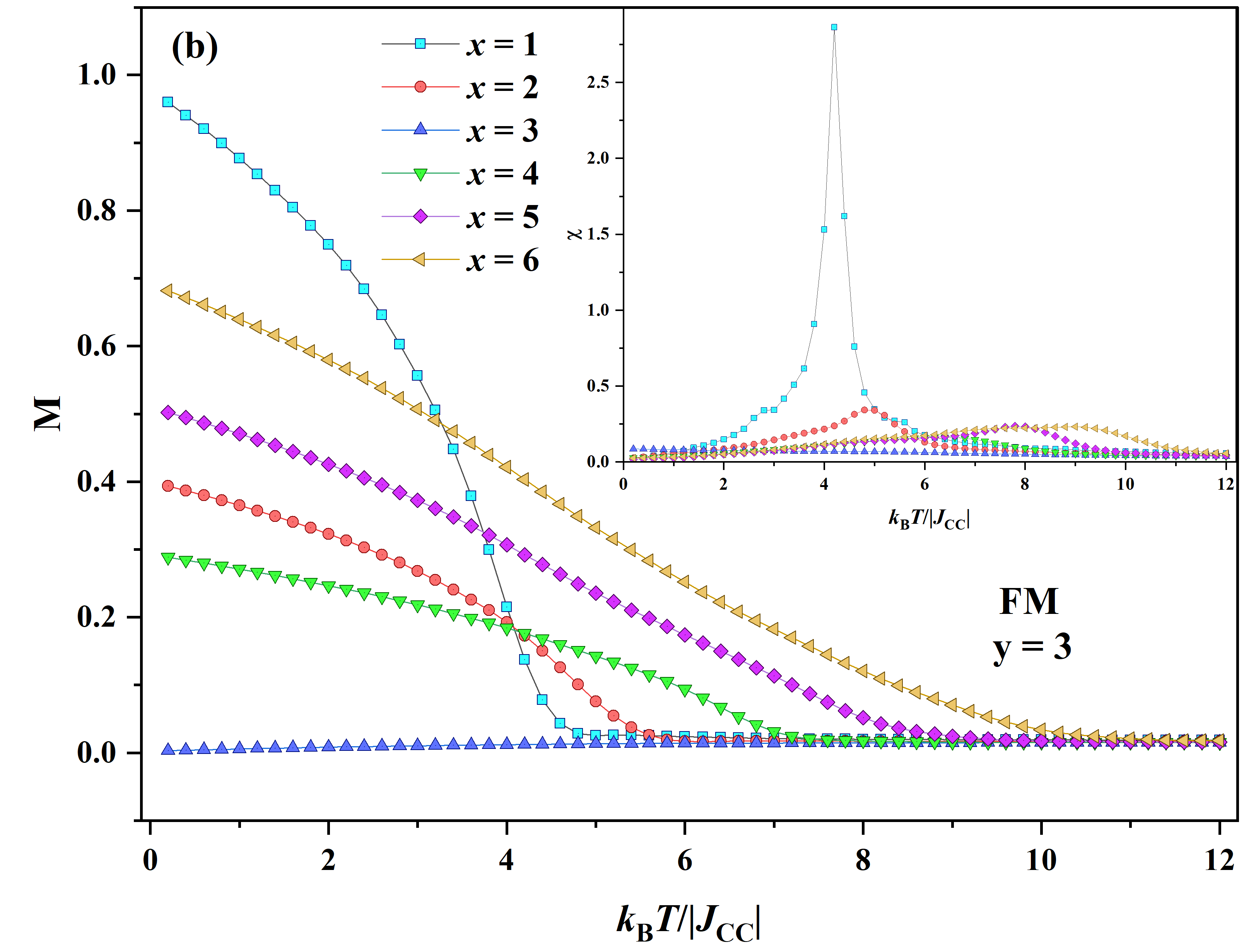}
    \caption{The plot of the dependence of magnetization on temperature under different \textit{f} values. The \textit{x} is equal to from 1 to 6 marked in the figure correspond to \textit{f} = 0.25, 0.40, 0.50, 0.57, 0.63 and 0.67, respectively. (a) the plot in the case of $J_{CF}$ $<$ 0. (b) the plot in the case of $J_{CF}$ $>$ 0.}
\end{figure}
The relationship between magnetization and temperature for various \textit{f}, as shown in Fig. 7(a). All saturation magnetizations in the AF superexchange coupling, are in accordance with the formula (4).  let $J_{FF}$ = 1.5$J_{CC}$, $J_{FC}$ = 0.5$J_{CC}$, \textit{p} = 0.5, and \textit{y} =3, we can calculate
\begin{scriptsize}
\begin{equation}
    T_P=\frac{5\,\sqrt{6}\,\sqrt{393\,x+\sqrt{\left( 7\,x^2+99\,x+567 \right) \,\left( 7\,x^2+687\,x+63 \right)}+7\,x^2+315}}{48k_B}
\end{equation}
\end{scriptsize}
There are discrepancies in phase transition temperature calculated between Heisenberg model and MFT under the same parameters. Thus the phase transition temperature calculated by MFT requires a series of transformations to apply to the Heisenberg model simulated by Monte Carlo method. The phase transition temperature calculated using MFT is well known to be proportional to the exchange constant. The same is true for the Heisenberg model [19]. Consequently, the phase transition temperature determined by MFT is proportional to that determined by Heisenberg model. Due to the ignorance of thermal disturbance by MFT, the phase transition temperature calculated by MFT is much higher than that of the Heisenberg model. We set a scale factor $\xi$ for conversion, \textit{i.e.}, $T_{P}^{H}=\xi T_P$. $\xi$ is independent of temperature but depends on the parameters of critical size \textit{L}, proportion \textit{p} and spin value \textit{S}. We set $\xi$(\textit{L}) = $\varepsilon \left( L \right) \frac{S_F\left( S_F+1 \right) \left( 1-p \right) +S_C\left( S_C+1 \right) p}{2}
$, the accuracy of the model is validated by fitting method. Through fitting the simulation data under the conditions that \textit{y} = 3, $J_{FF}$= -1.5, $J_{CC}$ = -1, $J_{FC}$ = -0.5, and \textit{p} = 0.5, the calculated the value of $\xi$ is 0.202 and $\varepsilon$ is 0.032, at \textit{L} = 20, as see in Fig. 8. \par
\begin{figure}[H]
    \centering
    \includegraphics[scale=0.31]{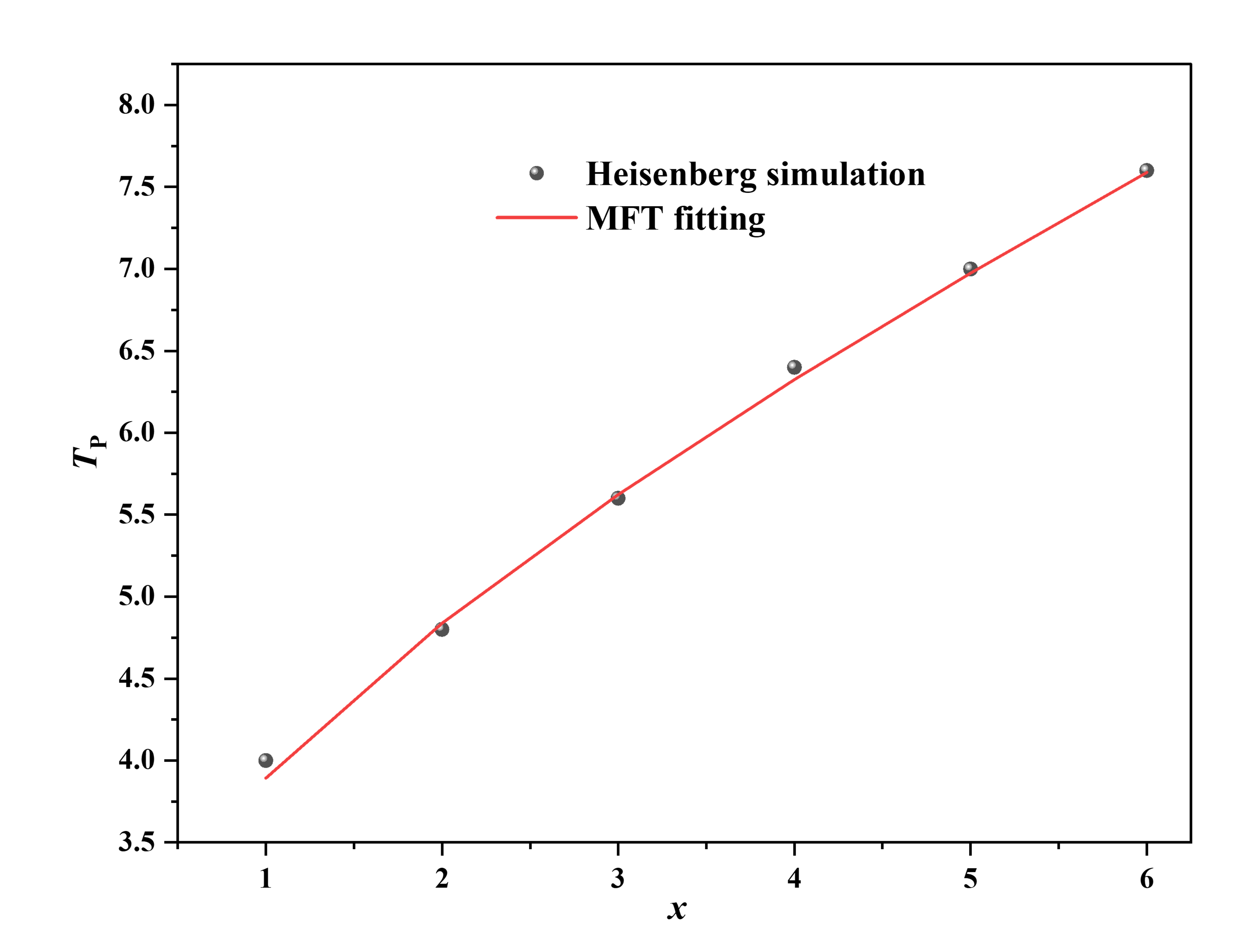}
    \caption{Fitting results of phase transition temperature for Heisenberg model simulation.}
\end{figure}
Figure 7(b) shows the M-T curve under FM superexchange coupling in Fe-O-Cr. We discovered an intriguing phenomenon in Fig. 7(a) and (b), \textit{i.e.}, when \textit{x} = 3, the magnetization is close to zeros. We can deduce from equal (12) that the $M_S$ under \textit{x} = 3 is zeros. Indeed, both magnetization and susceptibility exhibit antiferromagnetic properties. Notably, \textit{x} = 3 and \textit{y} = 3 are the most frequently occurring coordination numbers in a completely random system. According to numerous reports, when Cr/Mn is doped with Fe perovskite, the coordination relationship between Cr/Mn and Fe follows a binomial distribution [4]. In this case, the mean value of the coordination number all be three, \textit{i.e.}, \textit{x} = \textit{y} = 3. However, no experiment has observed that the magnetization tends to zero. This subject will be discussed in greater detail below. Notably, when \textit{x} is greater than 4, the magnetization exhibits the long-range order phenomenon near the phase transition temperature. Magnetic susceptibility peaks become extremely broad. This could be caused by the presence of a high degree of clustering. When \textit{x} approaches six, partial Fe-Fe clusters dubbed Fe-rich, form. The phase transition temperature increases as the Fe-O-Fe interaction increases. Weak Fe-O-Cr interaction reduces the phase transition temperature in part of the Fe-O-Cr interaction. They are in competition with one another, leading to long-range order.\par
\begin{figure}[H]
    \centering
    \includegraphics[scale=0.31]{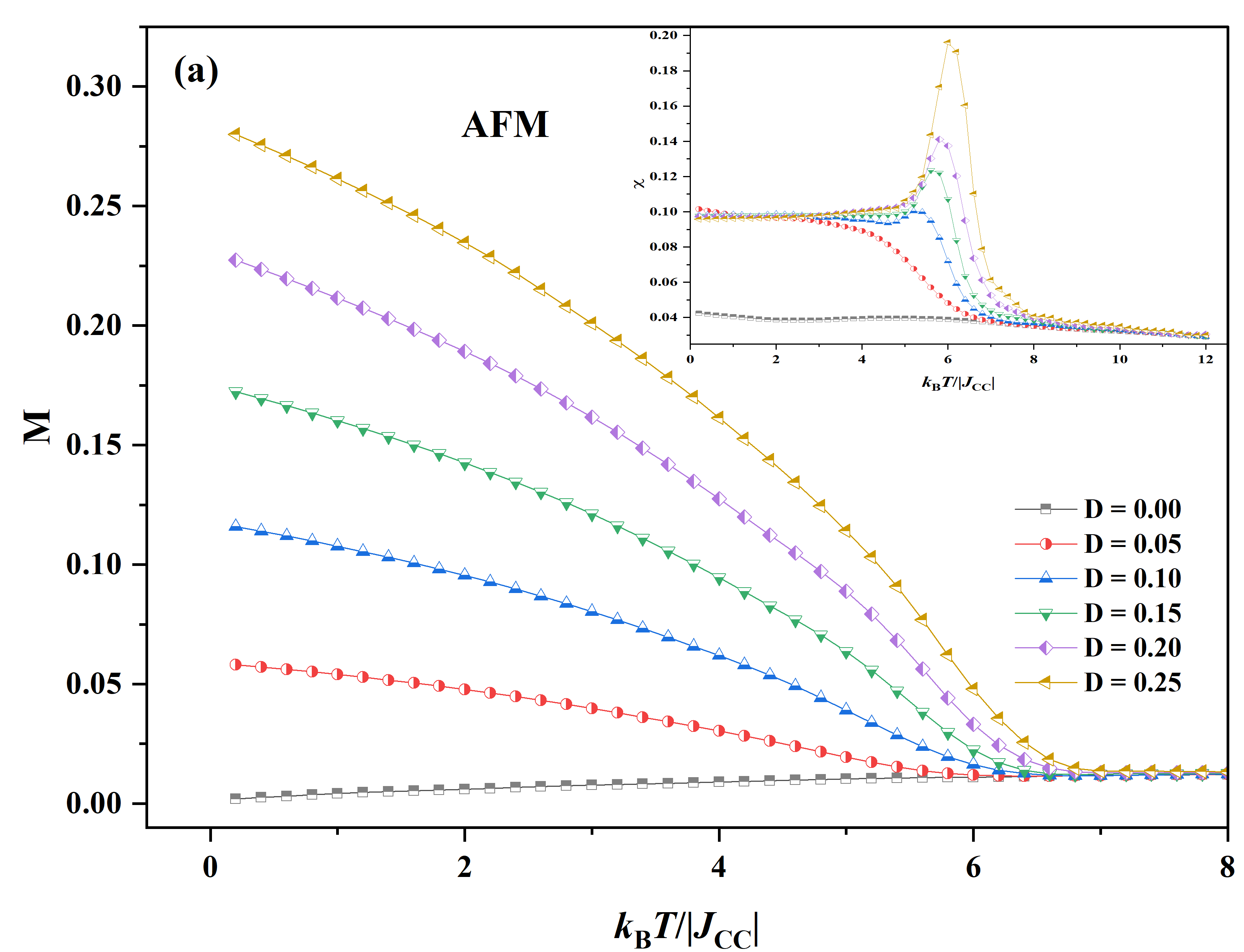}
    \includegraphics[scale=0.31]{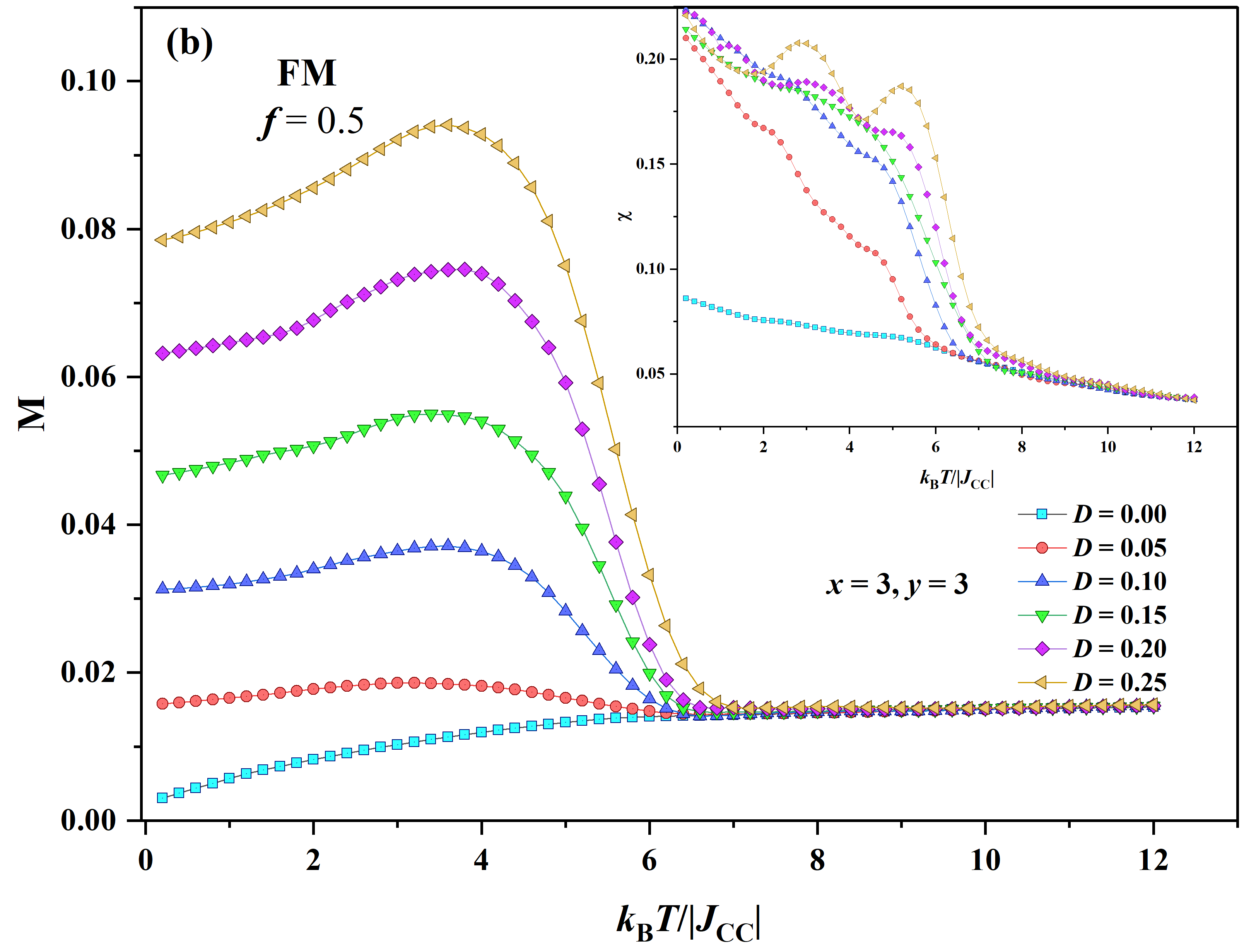}
    \caption{The plot of the dependence of magnetization on temperature under the interaction of different D. (a) in the case that AF superexchange interaction between Fe and Cr. (b) in the case that FM superexchange interaction between Fe and Cr. The inset is the dependence of magnetic susceptibility on temperature.}
\end{figure}
As previously stated, the magnetization is close to zero when \textit{x} = \textit{y} = 3. However, because of the DM interaction, the canting moment induces weak ferromagnetism (WFM) in the entire system, as illustrated in Fig. 9. This appears to account for the absence of experimental evidence for extremely weak saturation magnetization. Fe-O-Cr Antiferromagnetic coupling occurs between adjacent sites as a result of the Fe-O-Cr AF superexchange interaction. The net moment is increased as a result of the DM interaction. The inset of Figure 9(a) shows the susceptibility at various D values; the value of \textit{D} has little effect on the magnetic susceptibility peak position, indicating that the size of \textit{D} has little effect on the phase transition temperature.\par
\begin{figure}[H]
    \centering
    \includegraphics[scale=0.81]{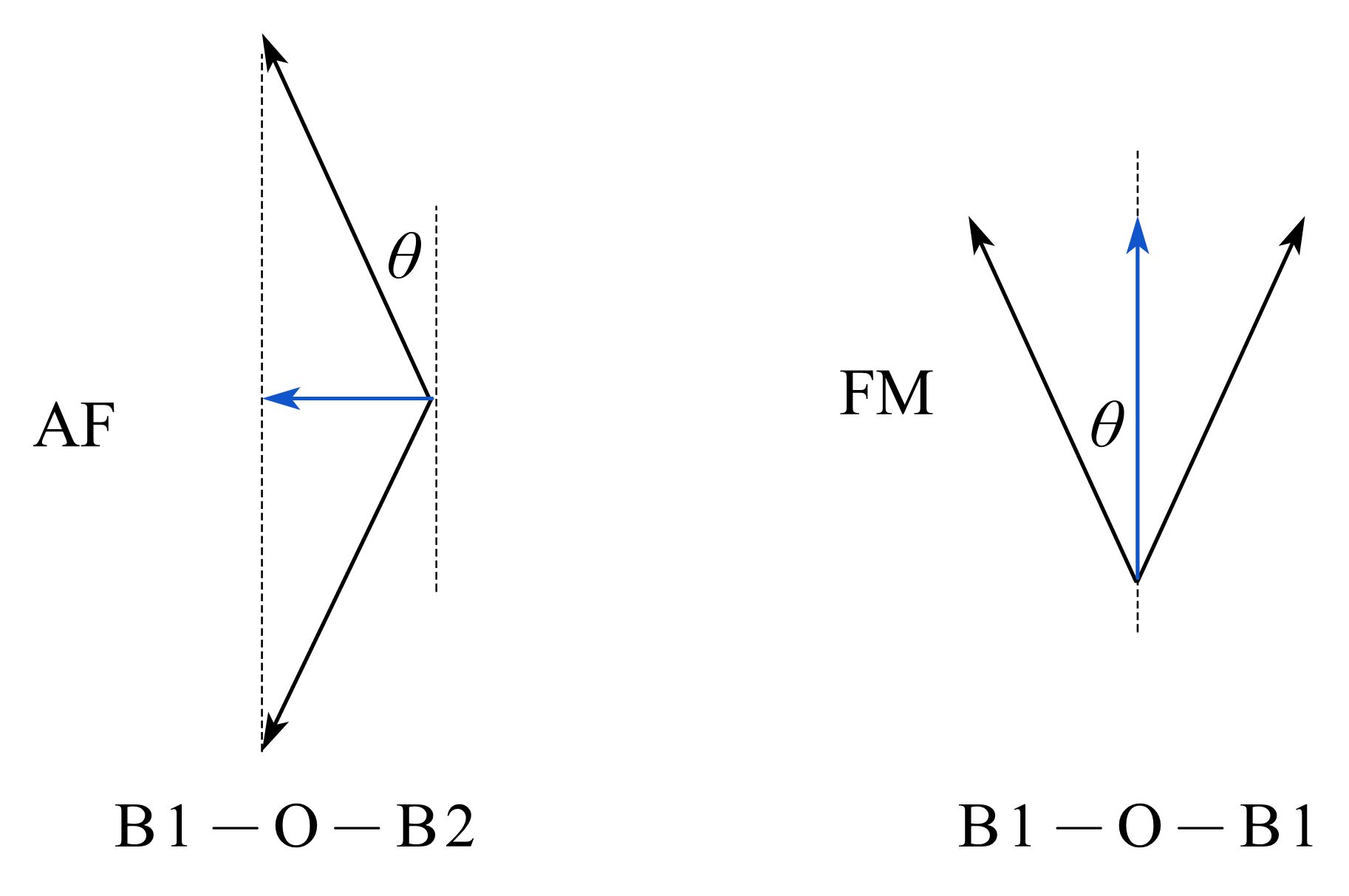}
    \caption{Discrepancy magnetic moment directions produced by DM interaction between AF and FM.}
\end{figure}
Indeed, when Fe-O-Cr is FM superexchange, the sublattice moments of F$^b$ and C$^a$ are in the same direction but are in opposition to those of F$^a$ and C$^b$. Marking that the sublattice of F$^b$ and C$^a$ is B1, and C$^b$ and F$^a$ are B2. When B1-O-B2 exists, the DM interaction generates a net moment in the same direction. When there is B1-O-B1 or B2-O-B2, however, the DM interaction reduces the magnetic moment, and points in the direction perpendicular to the net moment of B1-O-B2 as shown in Fig. 10. Furthermore, the net magnetic moment produced by this DM interaction is quite small, with a normal \textit{D} value of approximately 0.01 (in the order of magnitude in this paper). Thus, the effect of DM interaction, in this case, could be ignored. Moreover, there are two phase transition temperatures exhibited in the inset of Figure 9(b). The two humps may correspond to Neel temperature $T_N$ and Curie temperature $T_C$, respectively.\par
\begin{table}[H]
\setlength{\tabcolsep}{0.5mm}{
	\centering
	\caption{Based on the known values of \textit{D}, \textbf{$M_S$} and \textbf{$T_p$}, the values of \textit{x} and \textit{y} are inversed, where Set \textit{x} represents the set parameter and Pre \textit{x} represents the prediction parameter }
	\label{tab:2} 
	\begin{tabular}{ccccccccc}
		\hline\hline\noalign{\smallskip}
		\textbf{Num}&\textbf{\textit{D}} &\textbf{\textit{p}}&\textbf{$M_S$} &\textbf{$T_P$}&\textbf{Set \textit{x}}&\textbf{Set \textit{y}}&\textbf{Pre \textit{x}}&\textbf{Pre \textit{y}} \\
		\noalign{\smallskip}\hline\noalign{\smallskip}
		1&0.06&0.2&0.1349&11.0&4.3&5.8&4.29&5.80\\
        2&0.10&0.3&0.1317&9.1&5&4&4.97&3.85\\
        3&0.02&0.35&0.07495&10.3&5.8&4.6&5.82&4.65\\
        4&0.018&0.4&0.2832&5.4&1.5&4.2&1.52&4.21\\
        5&0.18&0.45&0.1611&9.6&5.0&5.5&4.91&5.70\\
\noalign{\smallskip}\hline
	\end{tabular}}
\end{table}
In order to verify whether the model can extrapolate the coordination distribution according to $M_S$ and $T_P$, we randomly simulate different parameters for \textit{x} and \textit{y}, and combined formula (10), (12), inverse the values of \textit{x} and \textit{y}. The results are listed in Table I. All the calculated results are in agreement with those of the expectation. We attempted to apply this theory to the study of the double perovskite phase transformation law and take the compound LaFe$_{0.5}$Cr$_{0.5}$O$_3$ into consideration. According to formula $T_N=2ZJS(S+1)/3k_B$, the $T_N$ of LaCrO$_3$ is approximately 256 K [20] and that of LaFeO$_3$ is 750 K [21]. Therefore, $J_{FF}$/$k_B$ and $J_{CC}/k_B$ can be calculated as 21.43 and 17.06 K, respectively. The phase transition temperatures vary in different reports. According to the majority of results, the phase transition temperature of LaFe$_{0.5}$Cr$_{0.5}$O$_3$ is above 300 K [22, 23]. Each of these results shows some of the same characteristics, \textit{i.e.}, low saturation magnetization. It is worth noting that the saturation magnetization of LaFeO$_3$/LaCrO$_3$ superlattice synthesized by Ueda K et al. is as high as 3 $\mu_B$ [24], which is very close to the theoretical value of 4 $\mu_B$. The Curie temperature is about 375 K, indicating a strong ferromagnetic exchange coupling. It demonstrates the ferromagnetic exchange of Fe-O-Cr within LaFeO$_3$/LaCrO$_3$ superlattice as a result of structural distortion [25, 26]. However, the conditions under which superlattice is prepared are extremely harsh. The bond angle of Fe-O-Cr tends to deviate from 180° when prepared normally. Pio Baettig et al. [27] investigated the phase transition temperature of double perovskite Bi$_2$FeCrO$_6$ using LDA+U method and predicted a phase transition temperature of approximately 130 K (corresponds to $J_{FC}/k_B$ = - 5.67 K calculated using MFT) in a relaxation structure with Fe-O-Cr bond angle less than 180$°$. They also predicted that the phase transition temperature of ideal double perovskite would not exceed 100 K. Although numerous experiments refute this conclusion, we believe it is credible. When Fe$^{3+}$ and Cr$^{3+}$ ions are arranged uniformly, the occurrence probability \textit{P} of the nearest number \textit{k} of Cr$^{3+}$ ions around one Fe$^{3+}$ ion (\textit{p} = 0.5) obey binomial distribution as
\begin{equation}
    P\left( k \right) =0.5^{k\left( 6-k \right)}·\frac{6!}{k!·\left( 6-k \right) !}
\end{equation}
The expected nearest number \textit{Ek} of Cr$^{3+}$ ions around one Fe$^{3+}$ ion is
\begin{equation}
    Ek=\sum_{k=1}^6{k·P\left( k \right)}=3
\end{equation}
$E_k$ satisfying the relation with \textit{x} and \textit{y} is
\begin{equation}
    Ek=\frac{6-x}{n_{F^a}}+\frac{6-y}{n_{F^b}}
\end{equation}
When the system is uniformly distributed, due to the symmetry of the sites, $n_{F^a}$=n$_{F^b}$ and \textit{x} = \textit{y}. According to Eq. 12, \textit{x} = \textit{y} = 3 can be determined.
We can calculate that the phase transition temperature in \textit{J}$_{FF}/k_B$ = -21.14 K, $J_{CC}/k_B$ = -17.06 K under uniform distribution (\textit{x} = \textit{y} = 3), is $\frac{\sqrt{525\,{J_{FC}^2}+58564}}{2}+$248.95 K, indicating that the phase transition temperature is at least 370 K. From this point, the discrepancy of ion radius between Cr and Fe, the disorder in the majority reaction process does not conform to the uniform distribution. LaFe$_{0.5}$Cr$_{0.5}$O$_3$ in the study of Paul Blessington Selvadurai A et al. [28] is proved to be disordered; the phase transition temperature is about 380 K according to our observation. We can reasonably assume that Fe and Cr are distributed uniformly in this system; $J_{FC}/k_B$ is calculated to be about 4.4 K, which corresponds to the $T_P$ of La$_2$FeCrO$_6$ of about 101 K and is very close to the conclusion of  Pio Baetting. According to the formula (10), different \textit{x} and \textit{y} values will correspond to different phase transformation temperature. We plot the phase transition temperature in Fig. 11 and observe that the phase transition temperatures are basically concentrated in the middle of the whole curved surface. In most cases, it should follow some statistical distribution, whereas it does not follow a perfect uniform distribution. The formula (110) is quite complex, but Fig. 11 shows a monotonous trend as a whole. Two extreme values are taken at \textit{x} = 0, \textit{y} = 0 and \textit{x} = 6, \textit{y} = 6, respectively. We calculate the two cases and the results are $5\sqrt{21}J_{FC}\,\,\mathrm{and} \max \left( 35J_{FF}, 15J_{CC} \right) \,\,$. This indicates that in a disordered system, $J_{FC}$ is less than $\frac{T_P}{5\sqrt{21}}$, which is used to distinguish the exchange constant of double perovskite.\par
\begin{figure}[H]
    \centering
    \includegraphics[scale=0.32]{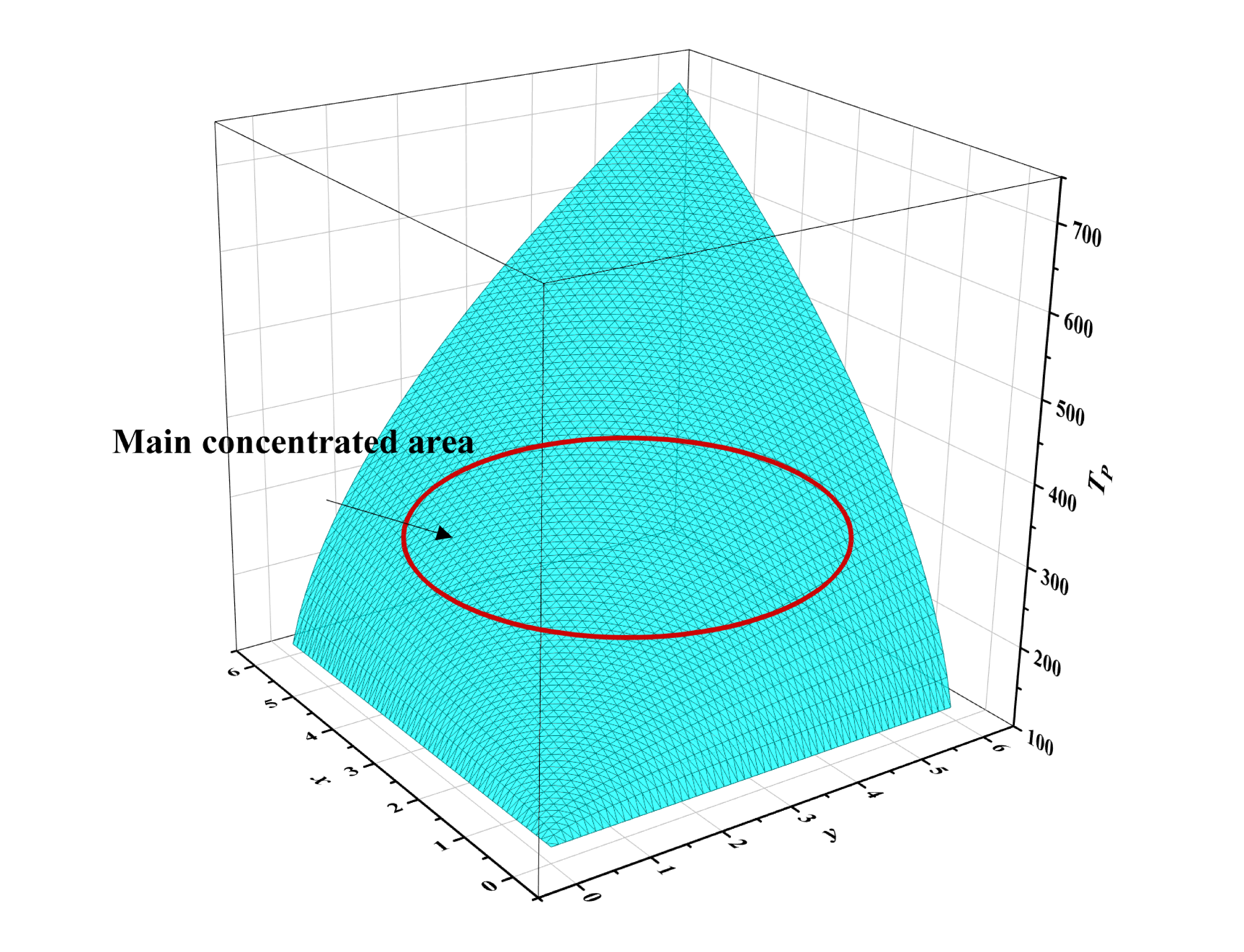}
    \caption{The plot of phase transition temperature $T_P$ surface corresponding to different \textit{x} and \textit{y} at \textit{p} = 0.5.}
\end{figure}
\section{Conclusion}
In conclusion, we have used the MFT algorithm and established a four-sublattice model suitable for the disordered system. By deducing the phase transition temperature and net magnetic moment equations, the coordination of a chaotic system can be calculated if $T_P$ and $M_S$ are known. We have tested it using the Monte Carlo method; the results are excellent and consistent with our expectations. We use Monte Carlo method to simulate the effect of different B-site distribution and DM interaction on magnetic properties. The magnetic properties are sensitive to the values of \textit{x} and \textit{y}. Larger \textit{x} and \textit{y} will trigger the long-range order of B-site magnetic ions. Additionally, our analysis reveals that the distribution of Fe / Cr in the majority of systems deviates from the uniform distribution, namely \textit{x} = \textit{y} = 3. In addition, we predict that the phase transition temperature of the ideal La$_2$FeCrO$_6$ double perovskite system is about 101 K based on this model. This model offers quantitative calculation method for B-site disorder system. It provides a new method for the quantitative calculation of magnetoelectric coupling in B site disordered system.\par
\section{DATA AVAILABILITY}
The data that support the findings ofthis study are available from thecorresponding author upon reasonablerequest.
\section{Competing Interests}
The Authors declare no Competing Financial or Non-Financial Interests.
\section{AUTHOR CONTRIBUTIONS}
The first author, Jiajun Mo, is responsible for programming, modeling, analysis and writing. Co-first author Min Liu is responsible for modeling and analysis, as well as result calculation,he has the same contribution as the first author. The third author, Shiyu Xu, is responsible for searching the literature, drawing and adjusting and revising the papers. The forth, fifth, sixth authors, Qinghang Zhang, Jiyu Shen and Puyue Xia put forward the improvement scheme for the paper respectively. First correspondent, Yanfang Xia, is the main director of this project. The second correspondent, Jizhou Jiang, provided technical support for the project.
\begin{acknowledgments}
The authors are grateful to Professors Andrew ThyeShen Wee and Alexandre Rykov for suggestion in this work. This research was supported partly by National Natural Science Foundation of China (grant number 12105137, 11447231), the Natural Science Foundation of Hunan Province, China (grant number  2020JJ4517), Research Foundation of Education Bureau of Hunan Province, China (grant number 19C1621, 19A434), the National Undergraduate Innovation and Entrepreneurship Training Program Support Projects of China (Grant No. 20200112).
\end{acknowledgments}

\end{document}